\documentclass[%
 reprint,
 superscriptaddress,
 amsmath,amssymb,
 aps,
]{revtex4-2}

\usepackage{graphicx}
\usepackage{dcolumn}
\usepackage{bm}
\usepackage{hyperref}
\usepackage[mathlines]{lineno}
\usepackage{physics}
\usepackage{siunitx}
\usepackage{xcolor}
\usepackage{subcaption}
\usepackage[export]{adjustbox}
\usepackage{easyReview}

\usepackage{comment}
\usepackage{txfonts}

\graphicspath{ {./figures/} }

\begin{document}

\preprint{APS/123-QED}

\title{Preservation of $^3\mkern-2mu$He ion polarization after laser-plasma acceleration}

\author{C.~Zheng}%
\email{chuan.zheng@hhu.de}
\affiliation{Peter Gr\"unberg Institute, Forschungszentrum J\"ulich GmbH, 52425 J\"ulich, Germany}%
\affiliation{Laser and Plasma Physics Institute, Heinrich Heine University D\"usseldorf, 40225 D\"usseldorf, Germany}%
\affiliation{ExtreMe Matter Institute EMMI, GSI Helmholtzzentrum f\"ur Schwerionenforschung, 64291 Darmstadt, Germany}%
\author{P.~Fedorets}%
\affiliation{Peter Gr\"unberg Institute, Forschungszentrum J\"ulich GmbH, 52425 J\"ulich, Germany}%
\author{R.~Engels}
\affiliation{Nuclear Physics Institute, Forschungszentrum J\"ulich GmbH, 52425 J\"ulich, Germany}%
\author{I.~Engin}%
\affiliation{Stabsstelle Arbeitssicherheit, Forschungszentrum J\"ulich GmbH, 52425 J\"ulich, Germany}%
\author{H.~Gl\"uckler}%
\affiliation{Central Institute for Engineering, Electronics and Analytics, Forschungszentrum J\"ulich GmbH, 52425 J\"ulich, Germany}%
\author{C.~Kannis}%
\affiliation{Laser and Plasma Physics Institute, Heinrich Heine University D\"usseldorf, 40225 D\"usseldorf, Germany}%
\affiliation{Nuclear Physics Institute, Forschungszentrum J\"ulich GmbH, 52425 J\"ulich, Germany}%
\author{N.~Schnitzler}%
\affiliation{Peter Gr\"unberg Institute, Forschungszentrum J\"ulich GmbH, 52425 J\"ulich, Germany}%
\author{H.~Soltner}%
\affiliation{Central Institute for Engineering, Electronics and Analytics, Forschungszentrum J\"ulich GmbH, 52425 J\"ulich, Germany}%
\author{Z.~Chitgar}%
\affiliation{Institute for Advanced Simulation, J\"ulich Supercomputing Centre, Forschungszentrum J\"ulich GmbH, 52425 J\"ulich, Germany}%
\author{P.~Gibbon}%
\affiliation{Institute for Advanced Simulation, J\"ulich Supercomputing Centre, Forschungszentrum J\"ulich GmbH, 52425 J\"ulich, Germany}%
\affiliation{Centre for Mathematical Plasma Astrophysics, Katholieke Universiteit Leuven, 3000 Leuven, Belgium}
\author{L.~Reichwein}%
\affiliation{Peter Gr\"unberg Institute, Forschungszentrum J\"ulich GmbH, 52425 J\"ulich, Germany}%
\affiliation{Institute for Theoretical Physics I, Heinrich Heine University D\"usseldorf, 40225 D\"usseldorf, Germany}%
\author{A.~Pukhov}%
\affiliation{Institute for Theoretical Physics I, Heinrich Heine University D\"usseldorf, 40225 D\"usseldorf, Germany}%
\author{B.~Zielbauer}%
\affiliation{Plasma Physics Department, GSI Helmholtzzentrum f\"ur Schwerionenforschung, 64291 Darmstadt, Germany}%
\author{M.~B\"uscher}
\email{m.buescher@fz-juelich.de}
\affiliation{Peter Gr\"unberg Institute, Forschungszentrum J\"ulich GmbH, 52425 J\"ulich, Germany}%
\affiliation{Laser and Plasma Physics Institute, Heinrich Heine University D\"usseldorf, 40225 D\"usseldorf, Germany}%
\date{\today}

\begin{abstract}

The preservation of nuclear spin alignment in plasmas is a prerequisite for important applications, such as energy production through polarized fusion or the acceleration of polarized particle beams. Although this conservation property has been the basis of numerous theoretical papers, it has never been experimentally confirmed. Here, we report on first experimental data from a polarized $^3\mkern-2mu$He target heated by a PW laser pulse, showing evidence for persistence of the nuclear polarization after acceleration to MeV energies. The finding also validates the concept of using pre-polarized targets for experiments at high-power laser facilities.

\end{abstract}

\maketitle


Spin dynamics in plasmas is an emerging field of research that studies how the spin of particles such as electrons, ions, and positrons evolves in plasma environments, especially under high-energy conditions such as in plasma accelerators or fusion reactors. Key aspects of  spin dynamics include generating spin polarization, understanding spin precession and the effects of intense electromagnetic fields in plasma environments. All potential applications critically depend on the ability to maintain the spin alignment of particles (i.e.\ their polarization) in the plasma for a sufficiently long period of time compared to the typical timescales involved  \cite{Thomas2020}.

The worldwide strategic processes \cite{CERN2022,Snowmass2022,Anderle2021} aiming at the realization of next generation’s particle accelerators point at the importance of polarized beams (leptons, protons, and ions) for fundamental research and various applications. Dozens of theoretical papers have been published in recent years on the feasibility of polarized beam generation in plasma accelerators (see e.g.\ Ref.~\cite{Buescher2020} for an overview).

Almost a century ago, M.\,Goldhaber inferred that fusion cross sections depend on the relative spin orientation of the participating nuclei (i.e.\ the 'fuel') \cite{Goldhaber1934}. Typical enhancement factors, e.g.\ for the total $d+t\rightarrow \alpha +  n$ cross section, amount to roughly 50\,\% \cite{Schieck2010,Baylor2023}. In their seminal 1982 paper \cite{Kulsrud1982}, Kulsrud et al.\ predicted time scales for polarization loss in a plasma of a magnetic fusion reactor to be much longer than the characteristic fuel burn-up period. A research project to measure the lifetime of spin-polarized fuel at the DIII-D reactor in San Diego is finally under way \cite{Baylor2023,Heidbrink2024}. Further advantages of polarized fusion are a facilitated neutron management and an increased tritium burn efficiency that may significantly reduce the startup tritium inventory \cite{Parisi2024}. Note that all these advantages should also apply to inertial fusion \cite{Temporal2012}; a more comprehensive discussion can be found in Refs.~\cite{Schieck2010,Ciullo2016}.

The typical association of nuclear spin alignment is with low temperatures, making it counter-intuitive that it could endure in a $10^8$ Kelvin plasma long enough to have practical applications. On the other hand, a theoretical study \cite{Thomas2020} of the scaling laws for the depolarization times indicated the feasibility of polarized particle acceleration in strong plasma fields. Experimental tests on the polarization conservation are  challenging \cite{Raab2014,Buescher2020,Baylor2023,Heidbrink2024} and have not been carried out up to now. 

Here, we report the findings of an experimental campaign using a nuclear-spin polarized $^3\mkern-2mu$He gas jet as a target \cite{Fedorets2022} for a Petawatt laser pulse. $^3\mkern-2mu$He -- the isospin partner of $^3\mkern-2mu$H -- has the advantage that it is not radioactive, can be polarized at room temperature, and stored over a long time in moderate (mT) magnetic holding fields. Laser-generated $^3\mkern-2mu$He plasmas, heated to the fusion resonance energies, are therefore an ideal testbed for polarized fusion. We present the first experimental evidence that the initial nuclear polarization is essentially preserved during the laser-induced heating and ionization and the subsequent acceleration of the ions from the plasma to MeV energies.

\begin{figure}[ht]%
\centering
\vspace*{3mm}
\includegraphics[width=1.0\linewidth]{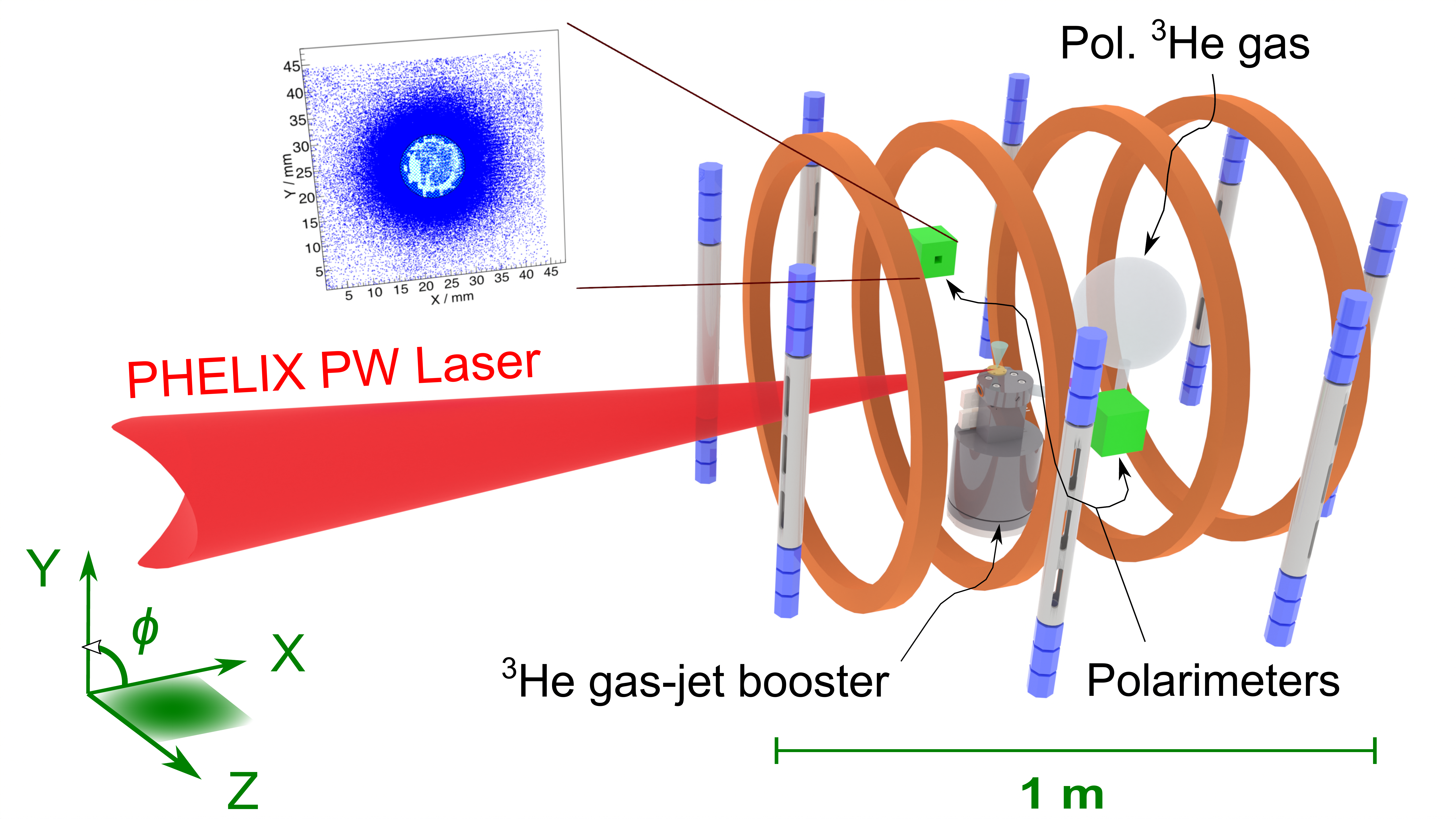}
\caption{Experimental setup at PHELIX: The laser beam (red) hits the polarized $^3\mkern-2mu$He gas jet that is emitted in  $+Y$ direction. The  magnetic system comprises a permanent magnet arrangement with eight vertical pillars (horizontal field in $+Z$ direction) and a Helmholtz-coil array for spin-orientation manipulation in the $X$-$Z$ plane. Two polarimeters (green), located inside radiation-protecting boxes (not shown), are mounted at right angles relative to the laser propagation axis $X$ with a direct view on the laser-plasma interaction region. The insert shows data from a detector plate in one polarimeter on elastically scattered $^3\mkern-2mu$He ions.} 
\label{fig1}
\end{figure}

\textit{Experiment}---The laser pulses were provided by the PHELIX Petawatt laser at GSI Darmstadt \cite{Phelix2010} and contained about 50\,J energy each. The optimal pulse duration of 2.2\,ps for the acceleration of $^3\mkern-2mu$He$^{1+,2+}$ ions  had been found during a previous measurement with unpolarized gas \cite{Engin2019}. The polarizer \cite{Mrozik2011} for the $^3\mkern-2mu$He gas was operated at Forschungszentrum J\"ulich, located roughly 250\,km from GSI. Unfortunately, due to small amounts of oxygen leaking into the polarizer, the initial gas polarization did not exceed $\sim$50\,\%, as opposed to values of $75$\,\% that were achieved during the preparatory phase \cite{Fedorets2022}. Due to various depolarization mechanisms, the initial gas polarization decreases exponentially in time. The polarized gas was stored in transport cells made of a special glass at an initial gas pressure of 3\,bar; a glass ball with fresh gas was re-supplied to the PHELIX target chamber (cf.\ Fig.~\ref{fig1}) every morning. With the help of a non-magnetic gas compressor this pressure can be enhanced before each laser shot to a maximum value of 30\,bar. The jet of 1\,mm diameter is formed with the help of a titanium de-Laval nozzle mounted on a non-magnetic valve, both located on top of the pressure booster. This leads to super-Gaussian distributed jet densities of a few times $10^{19}$\,cm$^{-3}$ which are required for the acceleration of the $^3\mkern-2mu$He ions to MeV energies \cite{Engin2019}. During the gas compression and jet formation the final $^3\mkern-2mu$He polarization decreased to half of the initial value \cite{Fedorets2022}.

Two identical polarimeters \cite{Zheng2022} are used to measure the remaining polarization of the accelerated $^3\mkern-2mu$He ions. They are placed at right angles relative to the laser propagation axis $X$ at distances of 95\,mm (at $+Z$, see Fig.~\ref{fig1}) and 264\,mm ($-Z$), respectively. The angular acceptances of the two polarimeters differ by a factor $\sim$8 --- below we present data for the near polarimeter, the results for its far counterpart support our findings but exhibit lower statistical accuracy. Each polarimeter contains a collimator (forming a 3\,mm-diameter beam), a secondary target for scattering of the ions, and five CR-39 detector plates (front, left, right, top, bottom) for particle identification.  Two types of reactions in the foil target are used for the analysis, i.e.\ Rutherford scattering of the accelerated $^3\mkern-2mu$He ions and the $^2\mkern-2mu$H($\vec{^3\mkern-2mu{\mathrm{He}}}$,\,$^4\mkern-2mu$He)$^1\mkern-2mu$H fusion reaction which transforms the $^3\mkern-2mu$He polarization information into a measurable azimuthal angular ($\phi$ as defined in Fig.~\ref{fig1}) asymmetry via a non-vanishing analyzing power $A$. 

The Rutherford scattering is dominated by few-MeV $^3\mkern-2mu$He ions \cite{Engin2019} deflected by the carbon atoms of the foil, which has a total thickness of $\sim$10$\,\muup$m (6\,$\muup$m CD$_2$ + 4\,$\muup$m CH$_2$). It is not sensitive to beam polarization ($A=0$), however, it can reveal asymmetries of the beam profile in the transverse phase space ($X$/$X^\prime$ and $Y$/$Y^\prime$) which are caused by the plasma acceleration and space-charge effects. The fusion reaction is much less sensitive to the transverse beam profile due to the large \textit{Q} value (18.35~MeV) and the small observed effects can be accounted for by the Rutherford-scattering data. The field of the Helmholtz coils causes a very small deflection of the $^3\mkern-2mu$He ions on their way to the polarimeter ($\sim 0.2^{\circ}$ for 2\,MeV ions) and does not affect the transverse beam profile. 

The initial $^3\mkern-2mu$He polarization along $+Z$ is maintained by a permanent magnetic holding field of 1.3\,mT \cite{Soltner2016}. Then the polarization ($P_Z/P = 1$) is (anti-)parallel to the momentum vector of the ejected ions (pointing at the two polarimeters) --- this is the case of longitudinal polarization, which is undetectable in the polarimeters. Additional Helmholtz coils (max.\ current 10\,A corresponding to 5\,mT) \cite{Fedorets2022,Soltner2016} allow one to rotate the initial $^3\mkern-2mu$He polarization in the horizontal plane towards $+X$ or $-X$ to achieve the transverse polarization case (max.\ $P_X/P = \pm 0.97$). Such a polarization can induce a top-bottom ($Y$ direction) rate asymmetry in scattering data of the polarimeter. One would expect no such asymmetry for unpolarized $^3\mkern-2mu$He gas which is precisely what was observed for the integrated top-bottom count-rate difference 
$\varepsilon=(N_{\mathrm t}-N_{\mathrm b})/(N_{\mathrm t}+N_{\mathrm b})=(-0.010\pm0.016)$. An irregular shape, limited to a small $\phi$-angle range, is observed in the corresponding Rutherford-scattering data on the front plate, and explains a count-rate fluctuation on the left plate.

\begin{figure}[ht!]%
\centering
\includegraphics[width=0.99\linewidth]{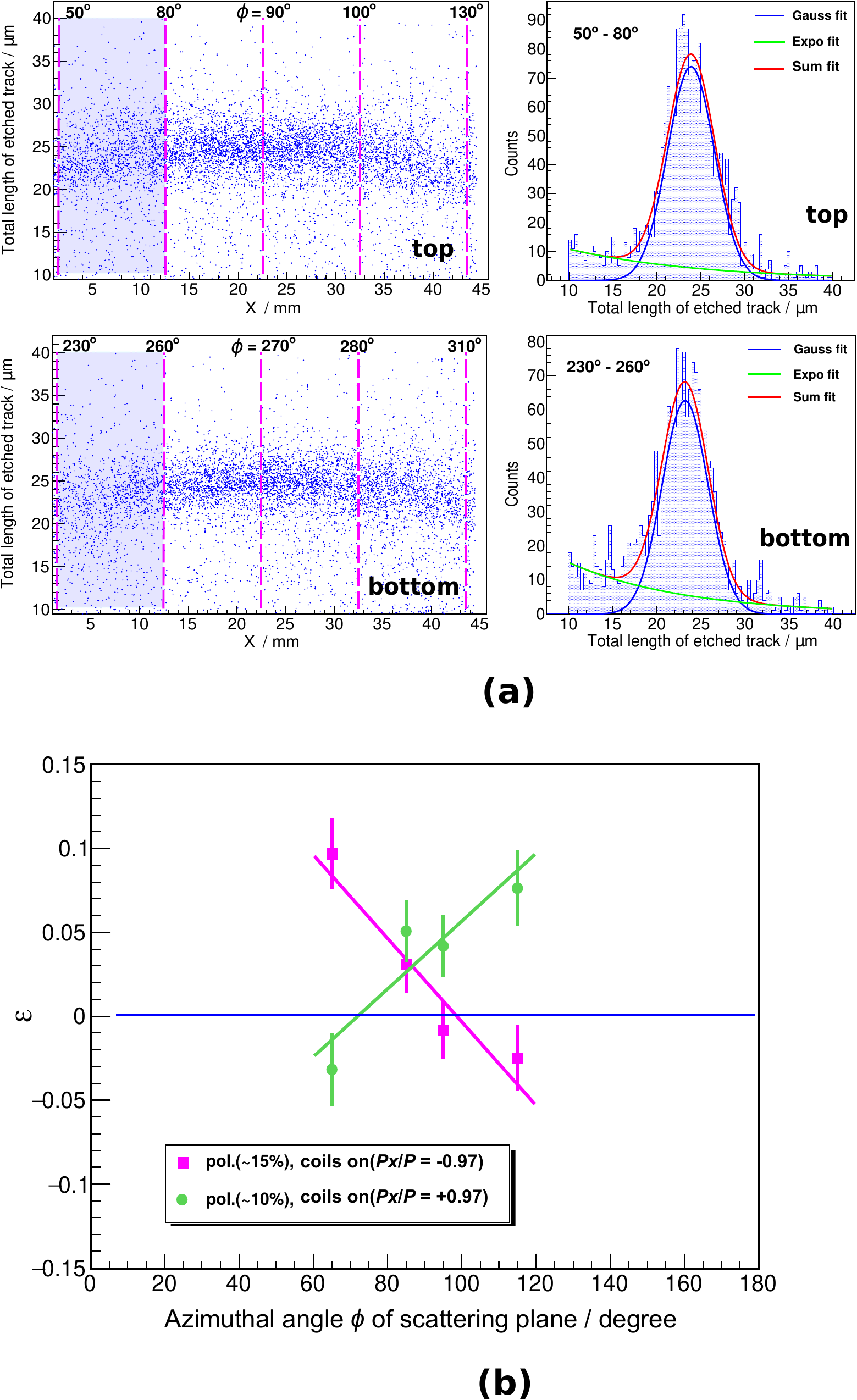}
\caption{(a) Hit pattern of the $\alpha$ particles on the top and bottom plates (left) for one field direction. Extraction of the fusion events by fitting a Gaussian function to the signal and an exponential function to the background (right). This procedure is applied in four angular bins in the $\phi$ range 50$^\circ$ to 130$^\circ$ (top detector) and on the opposite side. (b) Count-rate difference $\varepsilon$ of the $\alpha$ particles derived from the procedure described in (a). The error bar of each point is calculated from the total number of events under the signal peak including the background within a ``$3\,\sigma$ region'' of the Gaussian function. The green and pink lines are linear fits  to the data to provide visual guidance. The initial degree and orientation of the polarization in the jet are indicated.}
\label{fig2}
\end{figure}

We present the results of four days of data acquisition; during each day we recorded 5 to 6 laser shots for three different orientations of the $^3\mkern-2mu$He polarization and one for the control measurement with unpolarized $^3\mkern-2mu$He gas. We focus on a comparison of two data sets for the Helmholtz coils switched on but in opposite direction (transverse polarization), with emphasis on the azimuthal angular asymmetries of the $\alpha$ particles from the $^2\mkern-2mu$H($\vec{^3\mkern-2mu{\mathrm{He}}}$,\,$^4\mkern-2mu$He)$^1\mkern-2mu$H reaction. The angular distribution of the count-rate difference $\varepsilon(\phi)$ is depicted in Fig.~\ref{fig2}, which is deduced individually in four separate regions, since they are less affected by systematic uncertainties such as background suppression and detector efficiencies. The data points outside the 50$^\circ$ to 130$^\circ$ angular range could not be extracted from the left and right plates since the signal tracks on these plates were not sufficiently well formed during the etching procedure. 

For the transverse case a count-rate asymmetry must change sign for a reversal of the Helmholtz field orientation \cite{Raab2014,Zheng2022}. This is, of course, only the case under the assumption that the $^3\mkern-2mu$He ions are still transversely polarized after acceleration. Figure~\ref{fig2} shows that the measured angular distributions $\varepsilon(\phi)$ exactly reveal this ``spin-flip'' behavior. In the absence of a polarization persistence, the green and pink lines in Fig.~\ref{fig2} would both be horizontal (as indicated in blue). In contrast to the two data sets for transverse polarization, the $\varepsilon(\phi)$ distribution for the longitudinal case exhibits large statistical fluctuations and does not permit to make resilient statements on polarization conservation.

%
\textit{Discussion}---The observed asymmetries are larger than the naively expected $\varepsilon = P\cdot A<0.08$, where the initial gas polarization $P$ did not exceed 50\,\% and the analyzing power $|A|$ is smaller than 0.16 (for a maximum $^3\mkern-2mu$He ion energy of about 4\,MeV) \cite{Zheng2022, Engin2019}. One possible explanation is provided by simulations with the \textsc{geant4} code on the effect of asymmetric beam profiles in the polarimeter. These show that the count-rate asymmetries $\varepsilon(\phi)$ can be caused not only by the beam polarization but also by a beam profile with an irregular shape in the transverse phase space. Although the observed Rutherford scattering distributions exhibit only small fluctuations in the beam profiles, these can partially account for the observed $\phi$ asymmetries and thus mimic a higher polarization value.

In order to gain a better understanding of the prevalent acceleration mechanisms, we have conducted particle-in-cell (PIC) simulations with the \textsc{epoch} \cite{Arber2015} and the \textsc{vlpl} codes \cite{Pukhov1999,Pukhov2016} both with the inclusion of the T-BMT equation \cite{Bargmann1959_tbmt} describing spin precession. The assumed initial spin direction is that of laser propagation, i.e.\ $s_X = \hbar / 2$. When the laser pulse irradiates the $^3\mkern-2mu$He target, its ponderomotive force expels the electrons mainly towards the front and at right angles to laser propagation, leaving behind a plasma channel with a strong magnetic field (few $10^3$\,T). The charge separation then induces a Coulomb explosion, leading to the acceleration of the $^3\mkern-2mu$He ions at a $\pm 90^\circ$ angle. The same acceleration mechanism was  identified in the study of the unpolarized $^3\mkern-2mu$He and $^4\mkern-2mu$He targets \cite{Engin2019}. At the falling edge of the gas jet, a displacement between the electronic and ionic plasma components leads to an acceleration process in forward direction similar to Target Normal Sheath Acceleration \cite{willingale_prl06,lifschitz_njp2014}. These forward accelerated ions, however, make up only a small part of the total amount of accelerated $^3\mkern-2mu$He ions for our laser-target configuration. If shorter targets were utilized, more ions in forward direction would be expected \cite{Gibbon2022}. Both simulation codes show that the ions gain only small $s_Y, s_Z$ components in their spin vector due to a precession around the strong magnetic field of the  plasma channel which is fully compatible with the observed polarization conservation.

In simplified terms, the $^3\mkern-2mu$He$^{2+}$ ions perform two types of motion in the magnetic field of the plasma: the cyclotron motion and the precession of the nuclear magnetic moment $\vec{\mu}_\mathrm{i}$. This can be described by two non-relativistic equations as
\begin{align} 
\frac{\mathrm{d}\vec{v_\mathrm{i}}}{\mathrm{d}t} &= \frac{q_\mathrm{i}}{m_\mathrm{i}}( \vec{v_\mathrm{i}}\times\vec{B}) \label{eq:lorentz}\\ 
\frac{\mathrm{d}\vec{\mu_\mathrm{i}}}{\mathrm{d}t} &= \gamma_{\mathrm{h}}( \vec{\mu_\mathrm{i}}\times\vec{B})\,, \label{eq:spinrotion}
\end{align}
where $\vec{v_\mathrm{i}}$, $q_\mathrm{i}$ and $m_\mathrm{i}$ are ion velocity, charge and mass, respectively.  $\gamma_{\mathrm{h}}$ is the gyromagnetic ratio of $^3\mkern-2mu$He nuclei $\vec{\mu} = \gamma_{\mathrm{h}}\vec{s}$. The magnetic moment $\vec{\mu}$ is antiparallel to the nuclear spin $\vec{s}$, which is expressed by a minus sign of $\gamma_{\mathrm{h}}$.

We assume a simple model of the magnetic field $\vec{B}$ around the 1\,mm long plasma channel induced by the laser pulse with a focus spot of $10\,\muup\mathrm{m}\times10\,\muup\mathrm{m}$. The electrons are pushed forward along the channel, corresponding to an electric current $I_e$ (tens of kA) opposite to the direction of laser propagation, and according to Biot–Savart's law a vortex magnetic field $B_{\varphi}(r)= \mu_0 I_e/(2\pi r)$ builds up around the channel. This field then rotates the velocity vectors $\vec{v_\mathrm{i}}$ and the spin directions $\vec{\mu_\mathrm{i}}$ in the horizontal plane according to Eqs.(\ref{eq:lorentz}) and (\ref{eq:spinrotion}). The rotational effect on $\vec{v_\mathrm{i}}$ was measured in our previous experiment at $\pm 90^\circ$ for $^4\mkern-2mu$He ions \cite{Engin2019} and leads to a backward bending of $2^\circ$, which is also confirmed by the PIC simulations. 

Since the cyclotron motion and the spin precession are caused by the same local magnetic field $\vec{B}$, the rotational effect on the magnetic moment $\vec{\mu_\mathrm{i}}$ along a $^3\mkern-2mu$He ion trajectory can be deduced from the cyclotron and the precession frequencies of the $^3\mkern-2mu$He$^{2+}$ ions
\begin{align}
    \omega_{\mathrm{c}} = \frac{q_\mathrm{i} B}{m_{\mathrm{i}}} &= 0.641\times 10^{8}\,B\,[\mathrm{T}]\,\mathrm{rad/s} \\
    \omega_{\mathrm{s}} = \gamma_\mathrm{h}B  &= -2.038\times10^{8}\, B\,[\mathrm{T}] \, \mathrm{rad/s}
\end{align}
which have a fixed ratio of $\omega_\mathrm{s} / \omega_\mathrm{c} = -3.18$. Therefore, the precession angle is expected to be $8.5^\circ$ opposite to that of the cyclotron motion. This corresponds to a  small change of the $X$-component of the initial polarization from $P_X/P = 1$ to 0.99, which is consistent with the simulation result \cite{Gibbon2022}.

\textit{Conclusion}---We present results of a first experiment using a polarized target at a high-power laser facility and demonstrate their feasibility for the acceleration of polarized particles. Most notably we observe an angular asymmetry of the $\alpha$ particles in our polarimeter which we can only explain by a transverse polarization of the accelerated $^3\mkern-2mu$He ions. It is therefore shown that the transverse polarization of the nuclear spins is at least partially conserved (according to PIC simulations to more than 99\%) during plasma heating and acceleration to MeV energies.

An absolute polarization measurement of the $^3\mkern-2mu$He ions is not feasible because the values of the $^2\mkern-2mu$H($\vec{^3\mkern-2mu{\mathrm{He}}}$,\,$^4\mkern-2mu$He)$^1\mkern-2mu$H analyzing power are not known for the full energy range of the accelerated $^3\mkern-2mu$He ions, and the energy distributions of the $^3\mkern-2mu$He ions reaching the polarimeters are not precisely known. Our measurements were also limited by the low ($\lesssim 20$\,\% in the jet) initial polarization of the target gas due to vacuum leaks in our $^3\mkern-2mu$He polarizer. We note that our conclusions are based on relative measurements (like the change of sign of angular asymmetries in Fig.~\ref{fig2}b) and are averaged over 5--6 laser shots, making them robust to fluctuations of the laser intensity or target density, and plasma filamentation \cite{Engin2019}.

It is planned to continue the experiments at PHELIX at higher gas polarization, and using a narrower (0.5\,mm instead of 1.0\,mm) gas-jet target. This would have the advantage that the $^3\mkern-2mu$He ions should be dominantly emitted under $0^\circ$ and at significantly higher energies (10--15\,MeV) \cite{Gibbon2022}. In parallel we developed a polarized HCl gas target \cite{Huetzen2019} for laser- or beam-driven acceleration of polarized proton and electron beams \cite{Wen2019,Wu2019,Wu2020,Jin2020,Fan2022,Yan2022,Reichwein2022}. This target could also be operated with DI gas \cite{Sofikitis2018} and be combined with polarized $^3\mkern-2mu$He in order to further study options for polarized fusion with high-power lasers.

\begin{acknowledgments}
This work has been carried out in the framework of the JuSPARC (Jülich Short-Pulse Particle and Radiation Center) project \cite{Jusparc2020} and has been supported by the ATHENA consortium (Accelerator Technology HElmholtz iNfrAstructure) in the ARD programme (Accelerator Research and Development) of the Helmholtz Association. We acknowledge funding through the European Union's Horizon Europe research and innovation programme under grant agreement No. 101079773 (EuPRAXIA Preparatory Phase Project). Our results are based on experiment P191, which was performed at the PHELIX infrastructure at GSI Helmholtzzentrum f\"ur Schwerionenforschung, Darmstadt (Germany) in the context of FAIR Phase-0. The authors gratefully acknowledge the Gauss Centre for Supercomputing e.V. \cite{GaussCentre} for funding this project by providing computing time through the John von Neumann Institute for Computing (NIC) on the GCS Supercomputer JUWELS at J\"ulich Supercomputing Centre (JSC). Special thanks go to Werner Heil (retired Professor from Johannes Gutenberg-University Mainz) for providing the $^3\mkern-2mu$He polarizer and for answering many emergency calls.
\end{acknowledgments}



\bibliography{reference}

\end{document}